\documentclass[aps,showpacs,preprintnumbers,amsmath,amssymb, 
eqsecnum,
twocolumn, tightenlines
]{revtex4}

\usepackage[dvips]{graphicx}
\usepackage{bm}

\begin{document}

\bibliographystyle{apsrev} 

\title {Semiclassical solution for a Yang-Mills field with given energy}

\author{M. Yu. Kuchiev} \email[Email:]{kmy@phys.unsw.edu.au}

\affiliation{School of Physics, University of New South Wales, Sydney
  2052, Australia}

\affiliation{Institute of Advanced Studies, Massey University, Albany, New Zealand}

    \date{\today}

    \begin{abstract} 
A new classical solution for the Yang-Mills theory in which the Euclidean energy plays a role of a parameter is discussed. The instanton and sphaleron are shown to be particular examples of this more general solution. The energy parameter for them takes on special values, which are zero and sphaleron mass for the instanton and sphaleron, respectively. The solution is employed to describe the tunneling process, which is accompanied by a variation of the topological charge. A range of temperatures, where the new solution makes this tunneling more effective than the known mechanisms based on the instanton, caloron or sphaleron is found.
    \end{abstract}

    \pacs{11.15.-q, 
          11.15.Kc 
          }

    \maketitle


\section{Introduction}
\label{intro}
It is  known that variation of the topological charge of the gauge field can take place by overcoming an effective potential barrier. At zero temperature the most efficient mechanism is tunneling via the BPST instanton \cite{Belavin:1975fg}. 
For finite temperatures the tunneling effect is associated with the caloron
\cite{Harrington:1978ve,Harrington:1978ua}, which is a periodic instanton-type solution.
Recent developments and a corresponding list of references for the caloron can be found in reviews \cite{vanBaal:2009pn,Weinberg:2006rq}. An alternative approach to the periodic conditions was suggested in Ref. \cite{Khlebnikov:1991th}, where it was named the periodic instanton. It proved useful for applications related to multiparticle production and studies of the barion and lepton number violation in high energy collisions and early Universe  \cite{Bezrukov:2003qm,Bezrukov:2003er,Rubakov:1996vz}.
At finite temperatures there exists another possibility for variation of the topological charge, when the field goes over the top of the effective barrier, which separates valleys with different topological charges. In this case the field should be provided with sufficient amount of energy, which is obtained from the temperature. The configuration of the field at the top of the barrier is described by the sphaleron solution, as is explained by Klinkhamer and Manton \cite{Klinkhamer:1984di}.

Thus, the instanton and  caloron provide quantum tunneling through the barrier, whereas the sphaleron plays a role when temperature excitations, which are classical processes, are considered. However, in both these cases the probability of the transition is exponentially suppressed. For the instanton-caloron related processes it is suppressed as $\exp(-8\pi^2/g^2)$, where $g$ is the coupling constant. If the temperature $T$ is not extremely high, the sphaleron related probability is also suppressed, this time by the Boltzmann factor $\exp(-m_\text{sph}/T)$, where $m_\text{sph}$ is the sphaleron mass.

This work addresses a question of whether it is possible to enhance the probability of the transition, which is accompanied by variation of the topological charge. The idea is to combine the tunneling mechanism with thermal excitations. Assume that first the gauge field obtains (say, from the temperature) a fraction of energy that is needed to go over the top of the barrier, so that with this amount of energy a classical jump over the barrier remains impossible. However, the field can tunnel through the remaining part of the barrier. 
One can think that transition through this remaining part of the barrier may be 
more probable than penetration through the initial barrier, which takes place for instanton-related processes. We will see that indeed, this combined, excitation plus tunneling, mechanism proves effective.

This mechanism is described with the help of a particular configuration of the gauge field, in which the action reaches its minimum provided the Euclidean energy is fixed. 
Correspondingly, the energy plays the role of a parameter, which makes it convenient to refer to this field configuration as the energon \cite{energo}. The minimum of the action means that the energon describes a classical solution. We will discuss two points of view on the energon, looking at it either by fixing its energy, or the temperature. The former description may present interest for those applications, which take place in non-equilibrium states. The latter, which relies on the temperature, presumes that the equilibrium is attained.

Several properties of the energon match some of the properties of instanton, caloron, and sphaleron, though it remains distinctively different from other phenomena. The electric and magnetic fields of the energon are collinear, which is similar to the main property of the instanton, though  a coefficient of proportionality between the fields for the energon is more sophisticated, it is a  function of the Euclidean time. The instanton and sphaleron can be considered as particular examples of the energon for specific values of energy. When energy is zero, the energon is reduced to the instanton. If, on the other hand, the energy equals the sphaleron mass, the energon becomes identical to the sphaleron solution proposed in Ref. \cite{Ostrovsky:2002cg}. 
For negative Euclidean energies the energon exhibits a periodic behavior, which is reminiscent of a similar property of the caloron. If, however, energy is positive, the energon shows no periodicity. The energon has nonzero topological numbers, exhibiting therefore a nontrivial topological structure. However, in difference with the instanton and caloron the topological charge and Chern-Simons number for the energon are not integers, which can be compared with a non-integer Chern-Simons number  $1/2$ of the sphaleron. Discussing these and other properties of the energon below we will see that it represents an interesting and useful phenomenon.

\section{Instanton and sphaleron}
Consider the SU(2) non-Abelian gauge theory. Let us compare the instanton and sphaleron configurations of the  gauge field. The instanton solution for the potential $A_\mu^a$ and the field $F_{\mu\nu}^a$ reads ($\hbar=c=1$)
\begin{align}
&(A_\text{ins})_{\,\mu}^{\,a}~\,=~~~2~\eta_{\mu\nu}^a ~\frac{x_\nu}{x^2+\rho^2}~,
\label{Ains}
\\
&(F_\text{ins})_{\,\mu\nu}^{\,a}\,=\,-4\,\eta_{\mu\nu}^a \,\frac{\rho^2}{(x^2+\rho^2)^2}~.
\label{Finst}
\end{align}
Here $\rho$ defines the size of the instanton,
the conventional gauge is taken, and notation of the Euclidean formulation is adopted in which $x_\mu$ are the space-time coordinates in which the indexes run over $\mu,\nu, \,etc =1,\dots,4$, while the isotopic indexes for the SU((2) gauge group take on values $a,b,\, etc=1,2,3$.
The self-dual and anti-self-dual t'Hooft  symbols $\eta_{\mu\nu}^a$ and $\bar\eta_{\mu\nu}^a$ are defined conventionally
\begin{align}
&\eta_{\mu\nu}^a\,=\,\varepsilon_{a\mu\nu 4}+\delta_{\mu a}\delta_{\nu 4}-\delta_{\nu a}\delta_{\mu 4}~,
\label{eta}
\\
&\bar\eta_{\mu\nu}^a\,=\,\varepsilon_{a\mu\nu 4}-\delta_{\mu a}\delta_{\nu 4}+\delta_{\nu a}\delta_{\mu 4}~.
\label{aeta}
\end{align} 
Generically, the instanton is characterized by its size, location and orientation. To simplify notation only $\rho$ is presented explicitly in Eqs.(\ref{Ains}),(\ref{Ains}), the instanton position is set at the origin, while its orientation is specified indirectly, through the choice of the t'Hooft symbols. However, later on, see Subsection \ref{positive}, we will keep in mind that the instanton position and orientation are the free parameters.

Considering the sphaleron configuration let us remember that the sphaleron was introduced by Klinkhamer and Manton \cite{Klinkhamer:1984di} as a self-consistent solution for interacting gauge and Higgs fields in the context of the electroweak theory. Later on it was argued by Ostrovsky, Carter and Shuryak in \cite{Ostrovsky:2002cg} that important properties of the sphaleron structure can be described using the pure gauge field, which is restricted by specific conditions. This structure possesses only the magnetic field and has a simple analytical form. At the same time it retains the main features of the sphaleron solution, in particular its topological number (Chern-Simons number) $1/2$ is the same as in the sphaleron solution of \cite{Klinkhamer:1984di}. We will use the configuration 
found in \cite{Ostrovsky:2002cg} calling it the COS-sphaleron (an abbreviation from the names of the authors). The COS-sphaleron has the following form 
\begin{equation}
(A_\text{sph})^{\,a}_{\,m}\,=\,
\left(\delta_{am}- n_a n_m\right) \,\frac{f(r)}{r}+\epsilon_{ }\,\frac{n_n}{r}~.
\label{Asph}
\end{equation}
Here $x_m$ ($m,n,\,etc=1,2,3$) are spacial components of coordinates $x_\mu$, which comprise a 3-vector $\mathbf{r}$, $\mathbf{n}=\mathbf{r}/r$,
while $f(r)$ is the following function \cite{notation} 
\begin{equation}
f(r) \,=\, \frac{\rho^2-r^2}{\rho^2+r^2}~,
\label{phi}
\end{equation}
in which $\rho$ denotes the size of the sphaleron. 
In order to see similarity between the instanton and sphaleron solutions let us apply to the sphaleron in Eq.(\ref{Asph}) the gauge transformation defined by the unitary matrix
\begin{equation}
U\,=\,\exp\big[\,i  (\boldsymbol{\tau} \cdot \boldsymbol{n})\,\pi/4\,\big]
\label{U}
\end{equation}
in which $\tau^a$ are the Pauli matrices. 
The potential (\ref{Asph}) and gauge transformation identified in 
(\ref{U})
belong to a set of potentials defined in (\ref{Am}) and set of transformations (\ref{Ug}).
Applying Eqs. (\ref{alpha'})-(\ref{h'}), in which 
$\alpha=g=h=0$ and $\gamma=-\pi/2$ 
we rewrite the COS-sphaleron in the following form
\begin{equation}
(A_\text{sph})^{\,a}_{\,m}\,=\,2\,\varepsilon_{amn}\,\frac{x_n}{r^2+\rho^2}~.
\label{Asp}
\end{equation}
Compare now the instanton and sphaleron by matching Eq.(\ref{Ains}) with (\ref{Asp}). Take the same size $\rho$ in both cases. Clearly there are distinctions between the two configurations. The instanton in Eq.(\ref{Ains}) is obviously  a time-dependent solution, while the sphaleron in Eq.(\ref{Asp}) is a static configuration. But one can spot a similarity as well. By taking the instanton solution
at the moment of time $\tau=0$ one observes that its vector potential equals the potential of the sphaleron
\begin{equation}
{(A_\text{ins})^{\,a}_{\,m}}(x)_{\tau=0}\,=\,(A_\text{sph})^{\,a}_{\,m}(\mathbf{ r})~.
\label{A=A}
\end{equation}
This implies that the magnetic fields of these two configurations at this moment of time are also equal
\begin{equation}
{\mathbf{B}_\text{ins}^{a}}(x)_{\tau=0}\,=\,\mathbf{B}_\text{sph}^{a}(\mathbf{ r})~.
\label{B=B}
\end{equation}
Consider energy $\mathcal{E}^{(\text{E})}$ of the gauge field in Euclidean formulation
\begin{align}
&\mathcal{E}^{(\text{E})}\,=\,K+U^{(\text{E})}~,
\label{E}
\\
&K\,=~\frac{1}{2g^2}\,\int\,{\mathbf{E}^{a}}\cdot{\mathbf{E}^{a}}\,d^3r~,
\label{K}
\\
&U^{(\text{E})}\,=-\frac{1}{2g^2}\int\,{\mathbf{B}^{a}}\cdot{\mathbf{B}^{a}}\,d^3r~.
\label{pot}
\end{align}
Here the dot-product refers to conventional scalar product of spacial coordinates of vectors,
the electric field accounts for the kinetic energy $K$, while magnetic field is responsible for the potential energy $U^{(\text{E})}$ of the field. The latter is taken with the negative sign, which is specific for the Euclidean formulation. 
To mark the Euclidean formulation explicitly the upperscript $(\text{E})$ is used.  The same upperscript is applied below for such quantities as total energy, potential energy, Lagrange function etc. If necessary to refer to the Minkowski formulation, these quantities are either marked by the upperscript $(\text{M})$, or left without any upperscript,
for example the potential energy for the Minkowski formulation reads in this notation
$U^{(\text{M})}\,\equiv U\,=\,-U^{(\text{E})}$.
Consider behavior of the potential and kinetic energies as functions of the Euclidean time $\tau$ for the instanton solution.  From Eqs.(\ref{Finst}),(\ref{pot}) and (\ref{K}) one finds
\begin{equation}
-U_\text{ins}^{(\text{E})}(\tau)\,=\,K_\text{ins}(\tau)\,=\,\frac{3\pi^2}{g^2}\,\frac{ \rho^4}{(\tau^2+\rho^2)^{5/2}}~.
\label{Uinst}
\end{equation}
One verifies validity of this result by calculating the action
\begin{equation}
S_\text{ins}^{(\text{E})}\,=\,\int_{-\infty}^{\infty}(K_\text{ins}(\tau)-U_\text{ins}(\tau))\,d\tau=
\frac{8\pi^2}{g^2}~,
\label{action}
\end{equation}
which reproduces the well known instanton action. For the initial moment of time one finds from Eq.(\ref{Uinst})
\begin{equation}
-U_\text{ins}^{(\text{E})}(0)\,=\,\frac{3\pi^2}{g^2\rho}\,=\,m_\text{sph}~.
\label{msph}
\end{equation}
The last identity here follows from Eq.(\ref{B=B}), which  matches the magnetic field of the instanton with that of sphaleron, and from the simple fact that an absolute value of the potential energy for the sphaleron equals its mass. 
\begin{figure}[tbh]
\centering
\includegraphics[height=5.5 cm,keepaspectratio = true, 
]{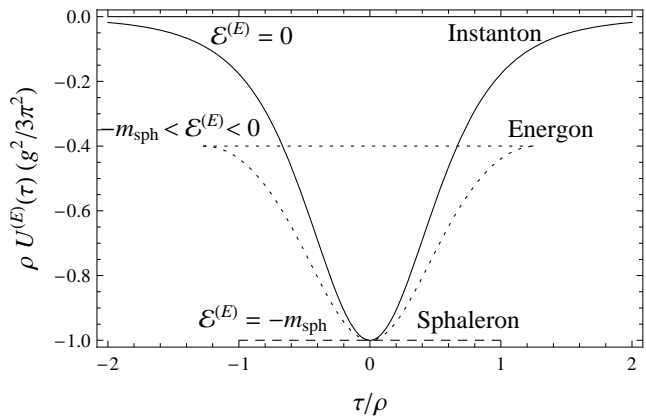}
\caption{
 \label{ras} 
The potential energy $U^{(\text{E})}(\tau)$ of the gauge field vs the Euclidean time $\tau$, energy is scaled by  $\rho g^2/3\pi^2$, time by $\rho$, Euclidean formulation in which the potential energy is negative is presumed. 
Solid horizontal line and solid curve - total energy and potential energy of the instanton (\ref{Uinst}), dashed line - energy of the sphaleron (\ref{msph}), dotted horizontal line and dotted curve - proposed energon configuration.
 }
 \end{figure}
\noindent
Remember that the sign minus in the left-hand sides of Eqs.(\ref{Uinst}),(\ref{msph}) is related to the Euclidean formulation, in which the motion within the instanton configuration is classically allowed, and the potential energy in Eq.(\ref{Uinst}) represents a potential well shown in Fig. \ref{ras}. The solid horizontal  line in this figure indicates the energy level of the instanton configuration, which according to Eq.(\ref{Uinst}) equals zero. 
The solid curve shows the profile of the potential energy for the instanton versus the Euclidean time. The minimum of this well is associated with the sphaleron configuration, which is static, has zero kinetic energy, and whose potential energy equals $-m_\text{sph}$.

In the Minkowski space instead of the potential well there exists an effective potential barrier $U^{(\text{M})}=-U^{(\text{E})}>0$.  In that case the instanton solution describes tunneling through the barrier, whose probability is $\propto \exp(-8\pi^2 /g^2)$. The sphaleron configuration describes events that take place at the top of the effective barrier, with the sphaleron mass equaling the energy of the effective barrier at its top. If the gauge field has a finite temperature, which is below the sphaleron mass,  $T<m_\text{sph}$, then the field can overcome the top of the barrier using thermal excitations, but a probability of this event is suppresses by the Boltzmann factor $\exp(-m_\text{sph}/T)$.

Thus, to overcome the barrier the field can either tunnel through it, or jump over its top, as has been well known previously. Consider the possibility of a new event, in which the gauge field first acquires  some energy from the temperature presuming that this energy is below the top of the barrier,  $0<\mathcal{E}^{(\text{M})}<m_\text{sph}$, and then tunnels through the potential barrier. The idea is that this `combined' mechanism can be more effective than either direct tunneling, or direct temperature excitation.

In terms of the Euclidean formulation our purpose can be illustrated using Fig. \ref{ras}, which shows two important energy levels, $\mathcal{E}_\text{ins}^{(\text{E})}=0$ for the instanton and $\mathcal{E}^{(\text{E})}=-m_\text{sph}$ for the sphaleron. Consider the situation when
the gauge field has an energy in between the two, $-m_\text{sph}\le \mathcal{E}^{(\text{E})}\le 0$. One such energy level is shown by the dotted horizontal line in Fig. \ref{ras}. One can try to adjust the configuration of the field 
in such a way, as to make a transition from the region described by negative $\tau$ to the 
region of positive $\tau$ the most effective. 
A possible configuration of the field, if exists, differs from the instanton simply because $\mathcal{E}_\text{ins}= 0 \ne \mathcal{E}^{(\text{E})}$,
and therefore the potential energy of this configuration should be different from the potential energy of the instanton. The dotted curve in the figure shows a possible profile of the potential energy. We presume that this possible configuration has a scaling factor, and that  by tuning this factor one can normalize the minimum of the potential curve for this configuration to make it equal to the minimum of the potential well for the instanton (minus sphaleron mass). 
The proposed potential well, which is  shown by the dotted line in Fig. \ref{ras}, in the
Minkowsi formulation represents a potential barrier.
The idea is to look for the configuration of the field, which provides the best possible chance (at the given energy) to overcome this effective barrier. If such configuration is found for any energy $-m_\text{sph}\le \mathcal{E}^{(\text{E})}\le 0$, it would
inevitably include the instanton and sphaleron configurations. 

\section{Energon}
\subsection{Variational description}
\label{energ}
Let us find a set of configurations for the gauge field, which interpolates between the instanton and sphaleron solutions presuming that the energy plays the role of a parameter. 
For simplicity in this Subsection we restrict ourselves to a simplified variational approach,
postponing a more rigorous treatment until the next Subsection \ref{minimum-action}.
We can  derive a possible useful form of the variational configuration from Eqs. (\ref{Ains}),((\ref{Asp}), which describe the instanton and sphaleron solutions. Introduce first $z_\mu$,
\begin{align}
&z_\mu\,=\,\big(\,x_1,x_2,x_3,q(\tau)\,\big)~,
\label{z}
\\
&z^2=\,x_1^2+x_2^2+x_3^3+q^2(\tau)~,
\end{align}
which is defined with the help of a  function $q\equiv q(\tau)$ (remember $\tau=x_4$) that should be found from a variational procedure formulated below. Define then the following potential
\begin{align}
&A^a_m(x)\,=\,2\,\eta^a_{m\nu}\,\frac{z_\nu}{z^2+\rho^2}~,
\label{Aam}
\\
&A^a_4(x)~=\,2 ~ \eta^a_{4\nu}~\frac{z_\nu}{z^2+\rho^2}~\dot{q}~,
\label{Aa4}
\end{align}
where $\dot{q}\equiv dq/d\tau $.
Calculating the field $F^a_{\mu\nu}$, which corresponds to this potential one observes 
that it has a simple appealing form
\begin{equation}
F^a_{\mu\nu}\,=\,-4\,\hat \eta^a_{\mu\nu}(\tau) \,\frac{\rho^4}{(z^2+\rho^2)^2}~.
\label{F}
\end{equation}
Here the modified $\hat \eta$-symbol is defined as follows
\begin{equation}
\hat \eta_{\mu\nu}^a(\tau)\,=\,\varepsilon_{a\mu\nu 4}+(\delta_{\mu a}\delta_{\nu 4}-\delta_{\nu a}\delta_{\mu 4})~\dot q(\tau)~.
\label{heta}
\end{equation}
For an arbitrary function $q(\tau)$ the modified symbol $\hat \eta_{\mu\nu}^a(\tau)$ 
differs from $\eta_{\mu\nu}^a$, in particular it is not self-dual. If, however, $q(\tau)= \tau$  then $\hat \eta_{\mu\nu}^a(\tau)=\eta_{\mu\nu}^a$,
similarly, $q(\tau)= -\tau$ leads to $\hat \eta_{\mu\nu}^a(\tau)=\bar \eta_{\mu\nu}^a$,
which makes the field selfdual or anti-selfdual correspondingly.
The field configuration described by Eqs.(\ref{Aam})-(\ref{heta}) exhibits similarities with the instanton solution from Eqs.(\ref{Ains})-(\ref{eta}). In particular, Eq.(\ref{F}),(\ref{heta}) show that its magnetic field equals magnetic field of the instanton, while its electric field differs from the instanton field by only a factor
\begin{align}
&\mathbf{B}^{a}(x)\,=\, {\mathbf{B}_\text{ins}^{a}}(z)~,
\label{BB}
\\
&\mathbf{E}^{a}(x)\,=\, {\mathbf{E}_\text{ins}^{a}}(z)~\dot q(\tau)~.
\label{EE}
\end{align}
Note though that the fields of the energon and instanton in these relations are taken at different values of Euclidean time, which equals $\tau$ for the energon and $q(\tau)$ for the instanton.
Since electric and magnetic fields of the instanton coincide, 
${\mathbf{E}_\text{ins}^{a}}={\mathbf{B}_\text{ins}^{a}}$,
Eqs.(\ref{BB}),(\ref{EE}) imply an interesting relation between the fields for the configuration defined by Eqs.(\ref{Aam})-(\ref{heta}) 
\begin{equation}
\mathbf{E}^{a}\,=~ \dot q~{\mathbf{B}^{a}}~.
\label{EB}
\end{equation}
Using Eq.(\ref{F}) one finds the density of the kinetic $k$ and potential  $u^{(\text{E})}$ energies
\begin{align}
&k~=~\frac{1}{2g^2}\,F^a_{m4}\,F^a_{m4}\,=\,
\frac{24\,\rho^4}{g^2}\,\frac{\dot{q}^2}{(z^2+\rho^2)^4}~,
\label{k}
\\
&u^{(\text{E})}\,=\,-\frac{1}{4g^2}\,F^a_{mn}\,F^a_{mn}
\,=\,-\frac{24\,\rho^4}{g^2}\,\frac{1}{(z^2+\rho^2)^4}~.
\label{u}
\end{align}
Integrating, one derives the kinetic and potential energies themselves 
\begin{align}
&K\quad\,=\,\int k ~d^3r~~~=~~~
\frac{3\pi^2\rho^4}{g^2}\,\frac{\dot{q}^2}{(q^2+\rho^2)^{5/2}}~,
\label{KK}
\\
&U^{(\text{E})}\,=\,\int u^{(\text{E})} \,d^3r\,=\,
-\frac{3\pi^2\rho^4}{g^2}\,\frac{1}{(q^2+\rho^2)^{5/2}}~.
\label{UU}
\end{align}
Let us find now the function $q(\tau)$ that provides a minimum for the classical action $S=\int L^{(\text{E})}d\tau$. This minimum is important for us since in Minkowski picture
it governs the probability of the tunneling through an effective potential barrier, see
Eq.(\ref{WE}) below. Specify first the Lagrange function of our system
\begin{equation}
L^{(\text{E})}\,=\,K-U^{(\text{E})}\,=\,
\frac{3\pi^2\rho^4}{g^2}\,\frac{\dot{q}^2+1}{(q^2+\rho^2)^{5/2}}~.
\label{L}
\end{equation}
Clearly, we can consider $q$ here as an effective coordinate within the Lagrange formalism, and $\dot q$ as a corresponding effective velocity. Realizing this, we can formulate and resolve the Lagrange equation for $q(\tau)$, finding hence the desired minimum for the action. Since we have only one function to specify, we can simplify our task by resolving from the very beginning the energy conservation law. Introduce the effective momentum $p$, which in the Lagrange formalism reads
\begin{equation}
p\,=\,\frac{\partial L^{(\text{E})}}{\partial \dot{q}}\,=\,
\frac{6\pi^2\rho^4}{g^2}\,\frac{\dot{q}}{(q^2+\rho^2)^{5/2}}~.
\label{p}
\end{equation}
Then derive the effective Hamiltonian $H^{(\text{E})}$, which corresponds 
to the given Lagrange function
\begin{equation}
H^{(\text{E})}\,=\,p\,\dot{q}-L^{(\text{E})}\,=\,
\frac{3\pi^2\rho^4}{g^2}\,\frac{\dot{q}^2-1}{(q^2+\rho^2)^{5/2}}~.
\label{H}
\end{equation}
Observe now that this {\em effective} Hamiltonian can be identified with the energy of the gauge field
$H^{(\text{E})}=K+U^{(\text{E})}$, where $K$ and $U^{(\text{E})}$ are defined in (\ref{KK}) and (\ref{UU}). This equality is not self-evident, for an arbitrary parametrization of the potential of the gauge field it does not hold.
The fact that it is satisfied supports the soundness of the parametrization of the potentials chosen in Eqs.(\ref{Aam}),(\ref{Aa4}) and makes arguments relying on the energy conservation law reliable. (Reasons explaining why the chosen parametrization of the potentials proves successful become clear in the next Subsection, which argues that this potential provides a local minimum for the action.) 

Stating the energy conservation law, $H^{(\text{E})}=\mathcal{E}^{(\text{E})}$, where $\mathcal{E}^{(\text{E})}$ is a constant, one finds from
Eq.(\ref{H})
\begin{equation}
\dot{q}\,=\,\pm\,\left(  1+\frac{g^2}{3\pi^2} \, \frac{\mathcal{E}^{(\text{E})} }{\rho^4} \, (q^2+\rho^2)^{5/2}\,\right)^{1/2}.
\label{dqdt}
\end{equation}
Here the sign depends on a choice of initial conditions. In what follows we will usually presume the sign plus. Substituting (\ref{dqdt}) in Eq.(\ref{p}) one expresses the momentum $p$ as a function of the coordinate $q$, which is necessary for calculation of the action, see (\ref{S}) below. Integrating Eq.(\ref{dqdt}) we derive 
\begin{equation}
\tau\,=\,\int^q\frac{dq}{\dot{q}}~.
\label{tq}
\end{equation}
Here $\dot q$ is a function of $q$ defined in (\ref{dqdt}), which makes the right-hand side a function of $q$. Resolving this function one finds $q(\tau)$ explicitly.

Consequently, Eq.(\ref{tq}) completely specifies the configuration of fields in Eqs.(\ref{Aam}),(\ref{Aa4}). We see that this configuration depends on two parameters, one being the scaling factor $\rho$, which is similar to the scaling factor in the instanton and sphaleron solutions. The other is the Euclidean energy, which
according to Eq.(\ref{dqdt}) takes on the values in the interval $-m_\text{sph}\le \mathcal{E}^{(\text{E})} < \infty$.  
Clearly, for nonpositive energies $-m_\text{sph} \le \mathcal{E}^{(\text{E})} \le 0$ $q(\tau)$ as a function of $\tau$ exhibits oscillations, whose properties are studied in some detail below, see Section \ref{negative}. For positive energies, $0<\mathcal{E}^{(\text{E})}$ the energon behaves differently. In that case a variation of  $\tau$ in some finite region  $-\tau_0/2\le \tau \le -\tau_0/2$  is accompanied by a corresponding monotonic variation of $q(\tau)$ from $-\infty$ to $\infty$. As a result the energon exhibits a phenomenon of localization in the finite region of $\tau$, see discussion in Section \ref{positive}.

Consider Eq.(\ref{Aam}) for the case when $\mathcal{E}^{(\text{E})}=0$. Then Eqs.(\ref{dqdt}),(\ref{tq}) show that $q(\tau)=\tau$, Eq.(\ref{z}) gives $z_\mu=x_\mu$. Consequently, Eqs.(\ref{Aam}),(\ref{Aa4}) are reduced to the instanton form (\ref{Ains}). This means that the instanton solution presents a particular example of the energon configuration, in which energy equals zero.
Equation (\ref{EB}) complies with this conclusion. It shows that if $q=\tau,~\dot q=1$, then electric and magnetic fields are equal, precisely what happens for the instanton solution. 

Take now the limit $\mathcal{E}^{(\text{E})}\rightarrow -m_\text{sph}$, where the sphaleron mass is defined in (\ref{msph}). Then according to Eq.(\ref{dqdt}) we have
$q,\,\dot q\rightarrow 0$, and Eqs.(\ref{KK}),(\ref{UU}) give $K\rightarrow 0$ and 
$U^{(\text{E})}\rightarrow -m_\text{sph}$. Thus, properties of the energon with $\mathcal{E}^{(\text{E})}= -m_\text{sph}$ are similar to the sphaleron configuration. 
Equation (\ref{EB}) supports this conclusion. It shows that for  $\dot q\rightarrow 0$ the electric field does not play a role, while the magnetic field of the energon equals the field of sphaleron, as evident from Eqs.(\ref{B=B}),(\ref{BB}) for $q\rightarrow 0$.

We see that the instanton and sphaleron can be considered as particular examples of the energon configuration.

\subsection{Minimum of classical action}
\label{minimum-action}
Consider the classical action and the energy of the gauge field
\begin{align}
&S^{(\text{E})}\,=\,\frac{1}{2g^2}\, \int d\tau
\int \! 
\big(\mathbf{E}^a\cdot \mathbf{E}^a+
\mathbf{B}^a\cdot \mathbf{B}^a\big)\,d^3r~,
\label{act}
\\
&\mathcal{E}^{(\text{E})}\,=\,\frac{1}{2g^2}\,
\int\!\big(\mathbf{E}^a\cdot \mathbf{E}^a-
\mathbf{B}^a\cdot \mathbf{B}^a\big)\,d^3r~.
\label{ene}
\end{align}
The signs in front of the terms $\propto \mathbf{B}^a\cdot \mathbf{B}^a$ here comply with the Euclidean formulation. Using conventional Cauchy's inequality we can write
\begin{equation}
\int \!\mathbf{E}^a\cdot \mathbf{E}^a d^3r\,\int \!\mathbf{B}^a\cdot \mathbf{B}^a
d^3r\,\ge\, \left(\int\! \mathbf{E}^a\cdot \mathbf{B}^a\,d^3r \right)^2\!.
\label{cauchy}
\end{equation}
Equations (\ref{ene}) and (\ref{cauchy}) allow one to estimate the action. Note firstly
that they imply
\begin{align}
\bigg(\int\! \big(\mathbf{E}^a\cdot \mathbf{E}^a 
& + \mathbf{B}^a\cdot \mathbf{B}^a\big)\,d^3r\bigg)^2  \,\ge\,
\label{E>}
\\
&\Big(2g^2 \mathcal{E}^{(\text{E})}\Big)^2
+4\left(\int\! \mathbf{E}^a\cdot \mathbf{B}^a\,d^3r\right)^2~.
\nonumber
\end{align}
From this inequality one derives the following restriction on the action
\begin{equation}
S^{(\text{E})}\ge\int\!\! d\tau \left[ \big(\mathcal{E}^{(\text{E})}\big)^2 
+\left(\frac{1}{g^2}  \int\! \mathbf{E}^a\cdot \mathbf{B}^a\,d^3r \right)^{\!2}\,\right]^{1/2}\!\!.
\label{estAc}
\end{equation}
The identity here can be reached only when identity in Eq.(\ref{cauchy}) is established. The latter takes place when
\begin{equation}
\mathbf{E}^a(x)\,=\,k(\tau)\,\mathbf{B}^a(x)~.
\label{EkB}
\end{equation}
Since the integrals in Eq.(\ref{cauchy}) run  only over spatial coordinates $d^3r$, the coefficient in Eq.(\ref{EkB}) is allowed to be an arbitrary (real-valued) function of $\tau$, $k=k(\tau)$. 

Equations (\ref{estAc}),(\ref{EkB}) show that if the energy of the gauge field $\mathcal{E}^{(\text{E})}$ is fixed, then one can search for a minimum of the action using condition (\ref{EkB}). The point is that  Eq. (\ref{EkB}) represents a non-linear differential equation for the 4-potential of the gauge field. Generally speaking, it may have presented a formidable problem. 

However, our previous discussion shows that the solution of this problem exists, Eq.(\ref{EkB}) can be resolved.  Compare (\ref{EkB}) with Eq.(\ref{EB}). See that these equations differ only by the name of a coefficient in them, so that changing  $k(\tau)$ in (\ref{EkB}), into $\dot q(\tau)$, $k(\tau)\rightarrow \dot q(\tau)$, one reduces (\ref{EkB}) to (\ref{EB}). Remember that the later equality was derived from the potential in (\ref{Aam}). We see that the potential (\ref{Aam}) gives a solution of (\ref{EkB}). 

This implies that the potential (\ref{Aam}) guarantees an equality in Eq.(\ref{estAc}).
Note a simple important point. Our search for a minimum of the action resulted in the particular potential (\ref{estAc}), which depends only on one function $q(\tau)$, so that if we minimize the action further by adjusting this function, we accurately represent the minimum. From the potential Eq.(\ref{Aam}) one derives the field (\ref{F}) and  calculates the spatial integral in  Eq.(\ref{estAc}) (remember this is an equality now). The result has a form
\begin{equation}
S^{(\text{E})}\,=\,\int L^{(\text{E})}\,d\tau~,
\label{SL}
\end{equation}
where $L^{(\text{E})}$ is the familiar Lagrange function, which was introduced in the previous Subsection in Eq.(\ref{L}). After that one minimizes the action in (\ref{SL}) by choosing the function $q(\tau)$ appropriately. Recall that precisely this procedure was undertaken in the previous Subsection \ref{energ} with the same Lagrange function, see Eqs.(\ref{L})-(\ref{dqdt}). Therefore we can state again that the function $q(\tau)$  is defined by Eq.(\ref{tq}). We conclude that the local minimum of the action, which is restricted by a given value of energy, is provided by the potential (\ref{Aam}), in which $q(\tau)$ is specified in Eq.(\ref{tq}). 

It is instructive to present this result in terms of the modified action, which takes into account the energy related restriction explicitly.
This can be achieved by using the conventional method of the Lagrange multipliers. 
Remember that along with the energy $\mathcal{E}^{(\text{E})}$ we also need to fix the size $\rho$ for the energon, which can be achieved if we introduce the following condition 
\begin{equation}
\min_\tau\,U^{(\text{E})}\,=\,-\frac{3\pi^2}{g^2}\,\frac{1}{\rho}~,
\label{max}
\end{equation}
where $U^{(\text{E})}$ is the potential energy defined in Eq.(\ref{pot}).
Adding a linear combination $\lambda(\tau)\mathcal{E}^{(\text{E})}+\sigma(\tau) U^{(\text{E})}$, where
$\lambda(\tau)$ and $\sigma(\tau)$ are the Lagrange multipliers, to the action $S^{(\text{E})}$ in (\ref{act}) we present the resulting modified action in the following form
\begin{equation}
S^{(\text{E})}_\text{res}=
\frac{1}{2g^2}\!
\int \! 
\Big(\epsilon(\tau)\,\mathbf{E}^a\cdot \mathbf{E}^a+
\frac{1}{\mu(\tau)}\,\mathbf{B}^a\cdot \mathbf{B}^a\Big)\,d^3r  d\tau\,.
\label{rest}
\end{equation}
Here restrictions posed by the fixed values of energy and scaling factor prompt a subscript `res' in the action, and the functions $\epsilon(\tau)$ and $1/\mu(\tau)$ include the terms from the original action in (\ref{act}) as well as the Lagrange multipliers, $\epsilon(\tau)=1+\lambda(\tau)$, $1/\mu(\tau)=1-\lambda(\tau)-\sigma(\tau)$.
Generally speaking, these functions should be derived from a complete minimizing procedure, but one can find without calculations that they are
\begin{equation}
\epsilon(\tau)\,=\,\mu(\tau)\,=\,\frac{1}{\dot q(\tau)}~.
\label{em}
\end{equation}
To verify this identity remember the similarity between the potentials for the energon (\ref{Aam}) and instanton (\ref{Ains}), which also manifests itself for their fields in Eqs. (\ref{BB}),(\ref{EE}). Basing on this resemblance one finds that the action for the energon can be derived from the conventional action for the gauge field in  Eq.(\ref{act}) using the following transformations in the action's integrand 
\begin{align}
&\tau \rightarrow q(\tau)~,\quad\quad\quad  d\tau \rightarrow \dot q(\tau)\,d\tau~,\\
&\mathbf{E}^a \rightarrow \mathbf{E}^a/\dot q(\tau), \quad \mathbf{B}^a \rightarrow \mathbf{B}^a~.
\label{teb}
\end{align}
These changes reproduce definitions in Eq.(\ref{em}) thus justifying the later identity. 
Alternatively, (\ref{em}) can be verified by writing the classical equations of motion that follow from it and comparing the result with Eqs.(\ref{BB}),(\ref{EE}).
Clearly, $\epsilon(\tau)$ and $\mu(\tau)$ play a role, which is  similar to the dielectric constant and permeability in conventional electrodynamics.

We learn from this discussion that the energon is characterized by two important properties. 
Firstly, it can be considered as a classical solution, in which Euclidean energy is fixed. 
Secondly, it can be viewed as a configuration, in which the electric and magnetic fields are collinear and the ratio of these fields is a nontrivial function of the Euclidean time.
Each of these two properties that the energon is not a selfdual configuration. 
To clarify this point let us take into account that the magnetic and electric fields of any selfdual solution are necessarily equal (their equality gives another representation of the selfdual condition).
Consequently, the kinetic energy, which is expressed via the electric field, and the Euclidean potential energy, which is derived from the magnetic field, are necessarily identical in magnitude and opposite in sign for any selfdual configuration. This makes the total energy of any selfdual solution zero. Thus, a selfdual solution is characterized by equal electric and magnetic fields and zero energy, in contrast to the energon which has dissimilar fields and nonzero energy. This discussion underlines the fact that the energon differs from any selfdual solution, including the instanton and caloron.

\section{Dynamic properties}
\label{basic}

\begin{figure}[t]
\centering
\includegraphics[height=5.5 cm,keepaspectratio = true, 
]{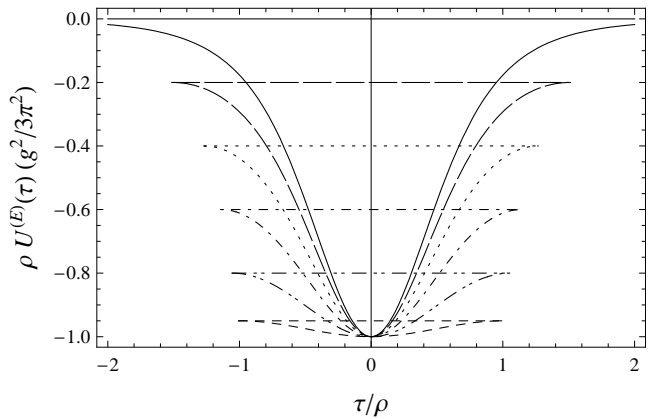}
\caption{
 \label{dva} 
The potential energy for the energon, units and assumptions same as in Fig. \ref{ras}. The horizontal lines from top to bottom, solid, long-dashed, dotted, dot-dashed, double-dot-dashed, and dashed - energy levels of the energon $\mathcal{E}^{(\text{E})}_\text{ener}=0,-0.2,-0.4,-0.6,-0.8,-0.95$ $\times 3\pi^2/(g^2\rho)$. The curves touching these levels - profiles of the potential energies $U^{(\text{E})}_\text{ener}(\tau)$  of the energon vs the Euclidean time $\tau$ for a given energy level.}
 \end{figure}
\noindent

\subsection{Negative Euclidean energies}
\label{negative}

To describe behavior of the energon at an arbitrary negative energy 
$\mathcal{E}^{ (\text{E}) }<0$
we calculate $q=q(\tau)$ numerically from Eq.(\ref{tq}), and then evaluate the potential energy from  Eq. (\ref{UU}). The results are shown in Fig. \ref{dva}.
Each curve in this figure represents a profile of the Euclidean potential energy 
$U^{(\text{ E})}(\tau)$ as a function of the Euclidean time $\tau$ for a fixed 
Euclidean energy  ${\mathcal E}^{(\text{ E})}$. A set of energy levels is shown by horizontal lines, each such line touches the corresponding potential curve. 
Observe that the potential curves are different for different energy levels ${\mathcal E}^{(\text{ E})}$, as one would expect.
However, if we keep the size-factor $\rho$  same, the potential curves for different energies have the same minimum $U^{(\text{ E})}(\tau=0)=-m_\text{sph}$. 
For each energy level the potential energy represents a potential well.
Hence, the system considered oscillates in this well. The period of 
oscillations is finite for all negative energies, while in the limit ${\mathcal E}^{(\text{ E})}\rightarrow 0$ the period diverges (see Eq. (\ref{t1/5}) below), and at 
${\mathcal E}^{(\text{ E})}= 0$ the motion is aperiodic.
When energy varies from 
${\mathcal E}^{(\text{ E})}=0$ to ${\mathcal E}^{(\text{ E})}=-m_\text{sph}$, the profile of the potential energy gradually shifts from the curve that describes the instanton solution to the minimum of the potential well, which is associated with the sphaleron. 

Consider in more detail the energon at extreme allowed values of its energy. As we know, the limit ${\mathcal E }^{(\text{E})}\rightarrow -m_\text{sph}$ implies that $q,\,\dot q\rightarrow 0$.
We can therefore expand the Lagrange function in Eq.(\ref{L}) as follows
\begin{equation}
L^{(\text{E})}\,\simeq\,\frac{3\pi^2}{g^2 \rho}\,
\left(\dot{q}^2-\frac{5}{2}\,\frac{q^2}{\rho^2}+1 \right)~.
\label{q=0}
\end{equation}
We derive from this that the variable $q(\tau)$ oscillates harmonically with frequency
\begin{equation}
\omega\,=\,\sqrt{\frac{5}{2}}\,\frac{1}{\rho}~.
\label{o}
\end{equation}
For half-period of these oscillations one finds 
\begin{equation}
\tau_{1/2}= (1/2)\,(2\pi/\omega)\approx 1.99\, \rho~,
\label{1.99}
\end{equation}
which complies with numerical data in Fig. \ref{dva} that shows that at low energies $\tau_{1/2}\simeq 2\, \rho$.

\begin{figure}[tbh]
\centering
\includegraphics[height=5.5 cm,keepaspectratio = true, 
]{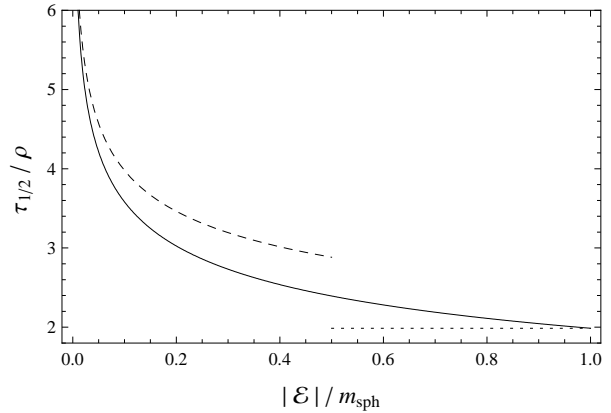}
\caption{
 \label{tri} 
Half-period $\tau_{1/2}$ for oscillations of $q(\tau)$ vs Euclidean energy $\mathcal{E}<0$, period in units of $\rho$, energy in sphaleron masses $m_\text{sph}$ from Eq.(\ref{msph}). Solid line - half-period calculated using Eq.(\ref{tq}), dotted line - asymptotic behavior described by Eq.(\ref{1.99}) near the sphaleron solution, dashed line - asymptotic form (\ref{t1/5}) near the instanton solution.}
 \end{figure}

Return now to the limit ${\mathcal E }^{(\text{E})}\rightarrow 0$, where the energon is close to the instanton solution. In this case Eq.(\ref{dqdt}) can be simplified 
$\dot q\simeq [1+g^2 \mathcal{E}^{ (\text{E}) }q^5 /( 3\pi^2\rho^4)]^{1/2}$. After that the integral in Eq.(\ref{tq}) is evaluated analytically
\begin{equation}
\tau\,\simeq\, _2F_1\left(\frac{1}{5},\frac{1}{2},\frac{6}{5},-\frac{g^2}{3\pi^2} \,\frac{q^5}{\rho^4} \,\mathcal{E}^{ (\text{E}) }\right)\,q~.
\label{E=0}
\end{equation}
Here $_2\!F_1(a,b,c,x)$ is the hypergeometric function.
Equation (\ref{E=0}) implies that 
for ${\mathcal E }^{(\text{E})}\rightarrow 0$ the half-period $\tau_{1/2}$ of the energon
oscillations (non-harmonic in this case) satisfies
\begin{equation}
\tau_{1/2}\simeq2.51\,q_\text{m}\,\simeq\,2.51\left(
-\frac{3\pi^2\rho^4}{g^2\mathcal{E}^{ (\text{E}) } }\right)^{\!1/5}
\!\!\!\propto \left| \mathcal{E}^{ (\text{E}) } \right|^{-1/5}\!\!.
\label{t1/5}
\end{equation} 
Here $q_\text{m}$ is the maximum for the classically allowed value for $q$, which is to be found from the condition $\dot q =0$, and the numerical coefficient comes from the hypergeometric function $2.51\approx  2\times_2\!\!F_1(1/5,\,1/2,\,6/5,1)$. Figure \ref{tri} presents half-period versus energy for all energies allowed.
\noindent

The found properties of energon at low negative energies, ${\mathcal E }^{(\text{E})}\rightarrow -0$, can be compared with the results of Khlebnikov, Rubakov, and Tinyakov \cite{Khlebnikov:1991th}. These authors developed an approach, calling it `periodic instantons', which is based on the idea that a classical solution can be approximated by a function periodic in $\tau$ that describes a chain of instantons and antiinstantons. The energon complies with this idea in the energy region considered, as evident from the discussion above, see the energy level shown by the dashed line in Fig. \ref{ras}. However, generically the energon is defined for a wide range of energies where it 
reveals very different properties, which may strongly deviate from the simple BPST instanton. To illustrate this point, recall the solutions that correspond to the energy levels at the bottom of the potential well in Fig. \ref{ras}, which are close to the COS-sphaleron. Note also that the energon exhibits an interesting phenomenon of the localization, which allows it to be defined in a finite interval of $\tau$, see Subsection \ref{positive}.

\subsection{Positive Euclidean energies}
\label{positive}
Consider now the region of positive Euclidean energies $\mathcal{E}^{ (\text{E}) }>0$. Eqs.(\ref{tq}),(\ref{dqdt}) in this case show that the function $q(\tau)$ is monotonic. Moreover it is easy to see  that it varies from $-\infty\le q\le \infty$ when $\tau$ runs over a finite interval  $0\le \tau\le \tau_0$, where $\tau_0$ 
is defined using  Eq.(\ref{tq}) as follows
\begin{equation}
\tau_0\,=\,\int_{-\infty}^{\infty}
\frac{dq}
{\dot q }\,=\,\rho \!\int_{-\infty}^{\infty}
\frac{dx}
{  \big(1+\epsilon\,(x^2+1)^{5/2} \,\big)^{1/2}}~.
\label{t0}
\end{equation}
Here $\dot q$ is taken from Eq.(\ref{tq}). To simplify notation it was presumed that $\dot q>0$, and that the lower limit of integration in Eq.(\ref{tq}) is set $\tau_0/2$. For positive energies the parameter $\epsilon=\rho g^2\,\mathcal{E}^{(\text{E})}/\,(3\pi^2)$, which is introduced in Eq.(\ref{t0}), is positive, $\epsilon>0$, which makes the integration convergent. Thus, Eq.(\ref{t0}) provides a sound definition for $\tau_0$. 

The above description of $q(\tau)$ is valid only within the interval $(0, \tau_0)$ where the function $q(\tau)$ is finite and allows one to define the energon. At the boundaries, $\tau\rightarrow 0,\,\tau_0$, this function diverges, $|q(\tau)|\rightarrow \infty$. As a result both electric and magnetic fields are decreasing here, $|E^a_m|,|B^a_m|\rightarrow 0$.  Thus the energon as a function of $\tau$  is well localized within the interval  $(0,\tau_0)$. This property will be called the `localization' of the energon.

The definition for the energon can be extended beyond the interval $(0,\tau_0)$, but  such extension is not uniquely defined and incorporates an interesting structure. Cover the entire axis $-\infty<\tau<\infty$ by a set of finite intervals of time $(\tau_k,\tau_{k+1})$, $k=0,\pm \,1,\dots$, $\tau_{-1}=0$. Presume then that in each given interval there exists an energon solution. We discussed above how to construct this solution for the interval  $(0,\tau_0)$. Similar approach is valid for any other interval $(\tau_k,\tau_{k+1})$. One takes the lower limit in the integral in Eq.(\ref{tq}) as $\tau_k$ thus specifying $q(\tau)$ in the interval $(\tau_k,\tau_{k+1})$, and then defines the energon using Eqs.(\ref{Aam}),(\ref{Aa4}).

The point is that one can use an arbitrary instanton solution in Eqs.(\ref{Aam}),(\ref{Aa4}). This means that the instanton orientation, size, and location, which are essential
parameters there, may be taken differently in different intervals $(\tau_k,\tau_{k+1})$. The opportunity for this divergence is related to the localization of the energon, which implies that what happens with the energon solution in the interval $(\tau_k,\tau_{k+1})$ does not interfere with the solution defined in any other interval $(\tau_{k'},\tau_{k'+1})$ because on the boundaries $\tau_k$ between the intervals the fields of the energon are all zero. There is one restriction in this construction though. Clearly, the substitution $\tau_0\rightarrow \tau_{k+1}-\tau_k$  makes Eq.(\ref{t0}) applicable to any interval $(\tau_k,\tau_{k+1})$. The resulting equation shows that the length 
$\tau_{k+1}-\tau_k$ of  the interval $(\tau_k,\tau_{k+1})$ is defined by the size of the instanton chosen for this interval. In other words the lengths of all intervals that cover the axis $\tau$ and the sizes of instantons used in (\ref{Aam}),(\ref{Aa4}) for each interval are interrelated. However, apart from this restriction all other instanton parameters in this construction are arbitrary.


We see that on the axis $-\infty<\tau<\infty$ the energon exhibits a rich structure governed by an infinite number of parameters that describe different instantons used to define the energon in different intervals of $\tau$. Moreover, even if all these parameters  are set equal, there still remains a freedom to choose the sign in Eq.(\ref{dqdt}), which may be taken differently in different intervals of $\tau$. 
Thus, on the axis $-\infty< \tau<\infty$ there exists an infinite variety of energon solutions. In particular, there is an option to construct solutions periodic in $\tau$, but they represent only a small fraction of all available opportunities. Generically the energons defined by this procedure do not show periodicity.

We learn from this discussion that the energon behaves very differently for different areas of energy. It is a uniquely defined periodic function of $\tau$ for negative energies. In contrast, for positive energies the energon  exhibits a phenomenon of the localization, which implies that it can be well described within a finite interval of $\tau$. Alternatively, the energon can be defined on the whole axis $\tau$, but the localization makes this definition dependent on an infinite set of parameters.

\section{Topological properties}
\label{topo}
For negative energies the energon configuration interpolates
between the instanton and sphaleron. Since these two later solutions are both topologically nontrivial, one has to expect that the energon possesses nontrivial topological properties as well. Consider the topological current 
\begin{equation}
K_\mu\,=\,\frac{1}{32\pi^2}\,\varepsilon_{\mu\nu\lambda\sigma}\,
\left( F^a_{\nu\lambda} A^a_\sigma-\frac{1}{3}\,\epsilon^{abc}\,
A^a_\nu A^b_\lambda A^c_\sigma
 \right)~.
\label{Kmu}
\end{equation}
(The signs comply with the Euclidean formulation.)
The divergence of this current equals the density of the topological charge
\begin{equation}
\partial_\mu K_\mu\,=\,\frac{1}{32\pi^2}\, F^a_{\mu\nu}\tilde F^{a}_{\mu\nu}~.
\label{dK}
\end{equation}
For the energon configuration (\ref{Aam})-(\ref{F}) we find the topological current 
\begin{equation}
K_\mu\,=\,\frac{1}{\pi^2}\,\frac{z^2+3\rho^2}{(z^2+\rho^2)^3}\,z_\mu~,
\label{ETC}
\end{equation}
and the corresponding density of the topological charge
\begin{equation}
\frac{1}{32\pi^2}\, F^a_{\mu\nu}\tilde F^{a}_{\mu\nu}\,=\,\frac{6}{\pi^2}\,
\frac{\rho^4\,\dot q}{(r^2+q^2+\rho^2)^4}~.
\label{Edens}
\end{equation}
Topological properties of the gauge field are closely related to the Chern-Simons number, which, up to a gauge transformation, equals  the integral of the charge density of the topological current
\begin{equation}
n\,=\int\,K_4\,d^3r-\nu~.
\label{nnu}
\end{equation}
Here the Maurer Cartan form (winding number) $\nu$ takes into account the fact that the topological current is not invariant under gauge transformations. 
Correspondingly the Chern-Simons number is also not invariant, its variation is given in Eqs.(\ref{CS'}),(\ref{CM}).
Using Eq.(\ref{ETC})  we find for the energon that the first term in (\ref{nnu}) reads
\begin{equation}
\int K_4\,d^3r\,=\,\frac{q\,(2\,q^2+3\,\rho^2)}
{4~(q^2+\rho^2)^{3/2}}~.
\label{CS}
\end{equation}
To assess the second term $\nu$ in Eq.(\ref{nnu}), we have to remember that the 
vector potential in Eq.(\ref{Aam}) exhibits a long tail $\propto 1/r$ at $r\rightarrow \infty$, which poses a problem. To eliminated it one has to choose the gauge appropriately. With this purpose let us make the gauge transformation defined by Eq.(\ref{Ug}), in which $\gamma=\gamma(r)$. Taking into account Eqs.(\ref{Am}),(\ref{alpha'})-(\ref{h'}) one verifies that in order to keep the potentials regular at the origin and at the same time eliminate the tail $\propto 1/r$ at infinity, it suffices to have
\begin{equation}
\gamma(0)\,=\,0~,\quad\quad\gamma(\infty)\,=\pi~,
\label{0infty}
\end{equation}
presuming that $\gamma(r)$ rapidly converges to its limiting value $\pi$ when $r\rightarrow \infty$.
According to Eq.(\ref{1/2})  condition (\ref{0infty}) implies that $\nu=-1/2$. Combining this result with Eqs.(\ref{CS}) and (\ref{nnu}) we derive the following expression for the Chern-Simons number
\begin{equation}
n\,=\,\frac{1}{2}\,\left(1+q~\frac{q^2+3\,\rho^2/2}{~(q^2+\rho^2)^{3/2}}\right)~.
\label{n}
\end{equation}
Since $q=q(\tau)$ is a function of $\tau$ the Chern-Simons number is a function of $\tau$ as well, $n=n(\tau)$. From Eq.(\ref{tq}) we know that $q(-\tau)=-q(\tau)$. Hence, the Chern-Simons number as a function of $\tau$ satisfies
\begin{equation}
n(\tau)+n(-\tau)\,=\,1~,
\label{nn1}
\end{equation}
which implies
\begin{equation}
n(0)\,=\,1/2~.
\label{n12}
\end{equation}
An additional insight into the properties of this function comes from a general relation between the Chern-Simons number and an integral over the density of the topological charge. Using Eqs.(\ref{dK}),(\ref{nnu}) one can present it in the following form
\begin{equation}
n(\tau)-n(0)\,=\,\frac{1}{32\pi^2}\,\int_0^{\tau}\! \!d\tau\int \!d^3r
\, F^a_{\mu\nu}\tilde F^{a}_{\mu\nu}~.
\label{nFF}
\end{equation}
This identity can be used to verify consistency of our results for the Chern-Simons number (\ref{n}) and the density of the topological charge (\ref{Edens}) for the energon. 
Alternatively, one can use it to derive anew Eq.(\ref{n}). (Though in this case one needs first to justify (\ref{n12}) independently. The latter goal may be achieved if a connection of $n(0)$ with the Chern-Simons number $1/2$ for the sphaleron is established, but here we will not press this argument further on.)

It is instructive to introduce the maximum and minimum of the Chern-Simons number, 
$0\le n_\text{min}\le n(\tau)\le n_\text{max}\le1$, which are achieved when $q$ in Eq.(\ref{n}) takes on the values $q=q_\text{m}$ and $q=-q_\text{m}$ respectively (remember $q_\text{m}$ is a maximum of $q(\tau)$ defined by condition $\dot q=0$, where $\dot q$ is given in Eq.(\ref{dqdt})). Equation (\ref{n}) shows that the two extrema satisfy
\begin{align}
&n_\text{min}+ n_\text{max}=1~,
\label{n+n=1}
\\
&0\le n_\text{min}\le 1/2\le n_\text{max}\le 1~.
\label{n<n}
\end{align}
Take the limit of low energy, $\mathcal{E}^{(\text{E})}\rightarrow 0$, when the energon is reduced to the instanton solution. Then
$q_\text{m}\rightarrow \infty$, and therefore we find
$n_\text{min,\,inst}=0$, $n_\text{max,\,inst}=1$, which shows that the total variation of the Chern-Simons number for the instanton,  $n_\text{max,\,inst}-n_\text{min,\,inst}=1$, reproduces the topological charge unity of the instanton, as it should. 
Take now the limit, when the energon is close to the sphaleron solution, 
$\mathcal{E}^{(\text{E})}\rightarrow  -m_\text{sph}$. Then, as we know, 
$q\rightarrow 0$ and therefore Eq.(\ref{n}) shows that $n_\text{sph}=1/2$, in agreement with results of Refs. \cite{Klinkhamer:1984di,Ostrovsky:2002cg} for the sphaleron. 

Generally speaking the found Chern-Simons number for the energon is non-integer. 
This is an interesting but not unique situation. Remember that the sphaleron has a non-integer  Chern-Simons number $1/2$. Remember also a family of configurations discussed in Ref. \cite{Ostrovsky:2002cg} in relation to the COS-sphaleron, which all have non-integer,
fractional Chern-Simons numbers.

\section{Tunneling through and thermal excitations over barrier}
Consider the action $S$ associated with the energon. The energy conservation
law (which is conveniently incorporated in our calculations, see discussion after Eq. (\ref{H})), makes the term $-\int H^{(\text{E})}d\tau $ in the action redundant. It suffices to take into account only the reduced, energy dependent part of the action, which reads 
$S( \,\mathcal{E}^{(\text{E})}\,)=\int p~dq$, where $p$ is a function of the coordinate $q$ specified by Eqs.(\ref{p}),(\ref{dqdt}). The action that describes the run of the system
from the stopping point $-q_\text{m}$ at the left wing of the potential well, see 
Fig. \ref{dva}, to the stopping point $q_\text{m}$ at its right wing 
(i. e. half of the total oscillation), reads 
\begin{equation}
S\big( \,\mathcal{E}^{(\text{E})}\,\big)\,=\,\int_{-q_\text{m}}^{\,q_\text{m}} p~dq~.
\label{S}
\end{equation}
Consequently, we find an estimate for the probability of the event, in which
the gauge field tunnels through the effective potential barrier, which exists in the Minkowski space
\begin{equation}
W\big( \,\mathcal{E}\,\big)\,\propto\,
\exp\big[-S\big( -\mathcal{E}\,\big)\,\big]~.
\label{WE}
\end{equation}
Here $\mathcal{E}\ge 0$ is the conventional positive energy of the gauge field in the Minkowski formulation, while $\mathcal{E}^{(\text{E})}=-\mathcal{E}$ is its negative Euclidean counterpart. To illustrate validity of Eq.(\ref{WE}) consider tunneling through the barrier for zero energy, $\mathcal{E}=0$. Then $q(\tau)=\tau$, and a simple calculation with the help of Eq.(\ref{p}) reveals that (\ref{WE}) reproduces the known instanton probability $W(0)=W_\text{ins}\propto \exp(-8\pi^2/g^2)$, as it should.

\begin{figure}[ht]
\centering
\includegraphics[height=5.5 cm,keepaspectratio = true, 
]{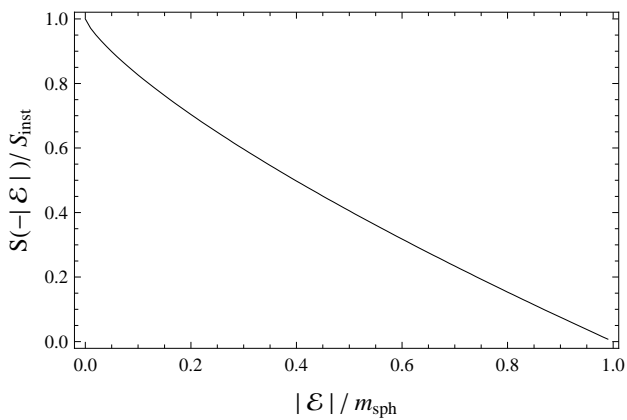}
\caption{
 \label{4etyre} 
The tunneling action of the energon $S(-|\mathcal{E}|)$ from Eq.(\ref{S}) vs the Euclidean energy of the gauge field $\mathcal{E}<0$, action in units of the instanton action $S_\text{ins}=8\pi^2/g^2$, energy in sphaleron masses $m_\text{sph}$ from Eq.(\ref{msph}).}
 \end{figure}

For arbitrary energy the action in Eq.(\ref{S}) can be calculated with the help of Eqs.(\ref{p}) and (\ref{dqdt}), which define $p$ as a function of $q$, $p=p(q)$. The results shown in Fig. \ref{4etyre} indicate that the action is decreasing for larger physical energies, which means an exponential increase for the tunneling probability (\ref{WE}). This complies with expectations based on a simple physical picture which presumes that the  higher the energy, the easier the tunneling should be. It also complies with the previous calculations presented in Fig. \ref{dva}, which show that for higher physical energy (lower Euclidean energy) the effective potential well in the Euclidean picture becomes smaller, and the corresponding effective barrier in the Minkowski formulation becomes smaller as well.

At a finite temperature $T$ the gauge field can populate any energy level $\mathcal{E} \ge 0$ with the probability, which is defined by the Boltzmann factor $\exp(-\mathcal{E}/T)$. The moment the field acquires the energy $\mathcal{E}$ the energon configuration becomes
allowed, which results in the probability of tunneling defined by Eq.(\ref{WE}). Therefore the probability of the combined  effect, when the field first acquires some energy from the temperature and then tunnels through the potential barrier is estimated as $W\propto \exp[-S(-\mathcal{E})- \mathcal{E}/T]$. To take this fact into account let us introduce the modified action $\tilde S$ as follows
\begin{equation}
\tilde S(T)\,=\,S(-\mathcal{E})+\mathcal{E}/T~,
\label{ST}
\end{equation}
which defines the probability of the combined excitation-tunneling at finite temperature
\begin{equation}
\tilde W(T)\,\propto\,\exp\big[-\tilde S(T)\,\big]~,
\label{WT}
\end{equation}
\noindent
To make the tunneling mechanism the most effective we are to minimize $\tilde S$ over the energy which implies
\begin{equation}
\frac{\partial \tilde S}{\partial \mathcal{E} }\,=\,
-S^\prime (-\mathcal{E})+\frac{1}{T}\,=\,0~.
\label{dS=0}
\end{equation}
This condition shows  that $\tilde S(T)$ is the Legendre transform of $S(-\mathcal{E})$.
In Eq.(\ref{dS=0}) the symbol $S^\prime $ indicates the derivative of the action over energy. When calculating this derivative via Eq.(\ref{S}) one needs to calculate the derivative of $p$ over energy, which according to Eqs.(\ref{p}),(\ref{dqdt}) equals
\begin{equation}
\frac{\partial p}{\partial \big(\mathcal{E}^{(\text{E})}\big)}\,=\,\frac{1}{\dot q}~~.
\label{dp}
\end{equation}
Using this fact we find from Eq.(\ref{S})
\begin{equation}
S^\prime (-\mathcal{E})\,=\,\int_{-q_\text{m}}^{q_\text{m}}\frac{dq}{\dot q}\,=\,\tau_{1/2}~.
\label{t1T}
\end{equation}
Here  the half-period  $\tau_{1/2}$ describes the time interval during which $q(\tau)$ goes from $-q_\text{m}$ to $q_\text{m}$. Eq.(\ref{t1T}) 
clarifies the physical meaning of
the condition (\ref{dS=0}). For a given temperature $T$ the energon that produces the most effective tunneling opportunity is the one, for which a half-period of oscillations of $q(\tau)$ equals the inversed temperature
\begin{equation}
\tau_{1/2}\,=\,1/T~.
\label{tau12}
\end{equation}
It is known that conditions of this type are typical when the field is exposed to a finite temperature, see e. g. \cite{LL9}. (There is a subtlety though. Conventionally, conditions for the field state that the full period of a periodic field equals the inverse temperature, while Eq.(\ref{tau12}) imposes a similar condition but for the half-period. The reason for this distinction is that conventional conditions are adopted for calculations related to quantities, which are at equilibrium at the given temperature, while here we consider 
a dynamical transition.)
\begin{figure}[tbh]
\centering
\includegraphics[height=5.5 cm,keepaspectratio = true, 
]{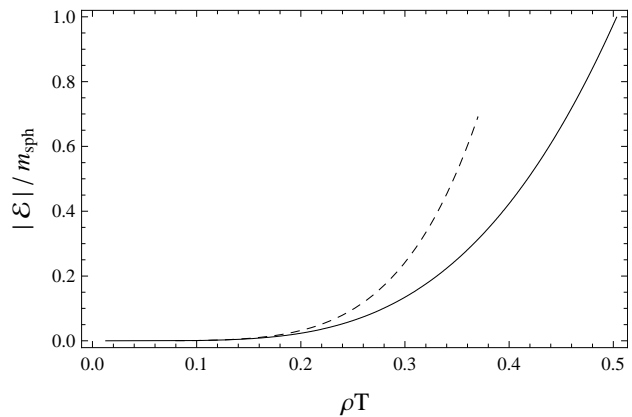}
\caption{\label{pat'} 
Euclidean energy $\mathcal{E}<0$ vs temperature $T$ specified by Eq.(\ref{tau12}). Energy in sphaleron masses, temperature scaled by the energon size $\rho$. Solid line - solution of (\ref{tau12}), dashed line - asymptotic form (\ref{ET5}).}
 \end{figure}
\noindent
Since we know $S(-\mathcal{E})$, see Fig.\ref{4etyre}, we can make the Legendre transform
required by Eq.(\ref{ST}) and find $\tilde S(T)$. First we solve Eq.(\ref{tau12}) finding energy as a function of temperature. The results shown by the solid line in Fig. \ref{pat'} comply with expectations based on simple physical arguments. Low temperatures obviously need low energy. The asymptotic form of this relation deduced from Eqs.(\ref{t1/5}),(\ref{tau12}) reads
\begin{equation}
|\mathcal{E}^{(E)}|/m_\text{sph} \,\simeq\,(2.51 \,\rho\, T)^5~.
\label{ET5}
\end{equation}
It is shown by the dashed lines in Fig. \ref{pat'}.
With increase of temperature the energy increases as well. 
The highest temperature for which Eq.(\ref{tau12}) admits solution is defined by the smallest possible half-period, which according to Eq.(\ref{1.99}) equals $(\tau_{1/2})_\text{min}\approx 2 \rho$. This value restricts the highest temperature $T_\text{max}\approx 1/(2\rho)$. The absence of solutions of Eq.(\ref{tau12}) for higher temperatures indicates that the direct process, in which the field acquires enough energy from the temperature to overcome the top of the barrier, presents the most efficient mechanism. Thus at  $\rho T>1/2$ the conventional COS sphaleron presents the most 
plausible opportunity for overcoming the barrier.
\begin{figure}[tbh]
\centering
\includegraphics[height=5.5 cm,keepaspectratio = true, 
]{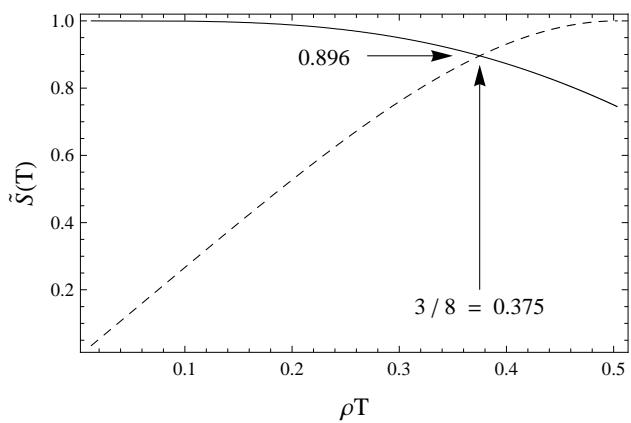}
\caption{
 \label{west} 
The energon action $\tilde S(T)$ from Eq.(\ref{ST}) vs temperature $T$. 
Solid line - comparison with direct tunneling, $\tilde S(T)/S_\text{ins}$, $S_\text{ins}=8\pi^2/g^2$ is the instanton action. Dashed line - comparison with direct thermal excitation over the top of the barrier, i. e. with sphaleron, $\tilde S(T)\,(T/m_\text{sph})$, $m_\text{sph}=3\pi^2/(g^2\rho)$ the sphaleron mass. Arrows indicate the point where the energon provides the most efficient mechanism, numbers give 
coordinates for this point. Temperature is scaled by the energon size $\rho$.}
 \end{figure}
\noindent

Figure \ref{west} shows the behavior of the energon action $\tilde S(T)$ in Eq.(\ref{WT}). Remember, the energon is a combined process, which uses both the tunneling mechanism and thermal excitations to overcome the barrier.  Figure \ref{west} compares it with two well known pure processes. One of them is the instanton, which represents the tunneling process and is governed by the tunneling exponent $S_\text{ins}=8\pi^2/g^2$. The solid line
in Fig. \ref{west} shows that $\tilde S(T)$ is always smaller than $S_\text{ins}$, thus indicating that at any finite temperature the energon is a more probable process than the pure tunneling. However, at low temperatures the difference between $\tilde S(T)$ and $S_\text{ins}$ is small, the solid curve converges to unity. With the increase of temperature this distinction becomes more prominent. 

The tunneling action for the caloron configuration equals $8\pi^2/\bar g^2$ \cite{Harrington:1978ua}, which is similar to the instanton action. The difference is that the coupling constant for the caloron depends on the temperature. However, this dependence is of a conventional logarithmic nature. Therefore it results only in a preexponential factor. The latter is important for detailed calculations, but if we are content with an exponential estimate, which is shown in Fig. \ref{west}, then the caloron and instanton give a similar contribution to the tunneling probability.

A dashed curve in Fig. \ref{west} presents a ratio
$T \tilde S(T) /m_\text{sph}$, which compares a capability  of the energon to overcome the barrier with the sphaleron, i. e. direct Boltzmann excitations over the top of the barrier (given by the sphaleron mass $m_\text{sph}=3\pi^2/(g^2\rho)$). At low temperatures the energon is more efficient, the mentioned ratio being smaller than 1. With an increase of temperature the distinction becomes less prominent and eventually disappears at $\rho T\approx 1/2$, where the dashed curve reaches unity. Above this temperature  the mechanism related to sphaleron, i. e. direct Boltzmann excitations, is the dominant process.

Observe a region in the vicinity of $\rho T= 3/8$, which is distinguished by arrows in Fig. \ref{west}.  Here the energon-based tunneling process prevails over both the instanton and sphaleron based mechanisms. At this temperature the tunneling action for instanton, $S_\text{ins}$, equals the factor $m_\text{sph}/T$, which governs the Boltzmann excitations. 
The energon action $\tilde S(T)$  at this point represents $0.896$ ($\approx 90 \% $) of 
$S_\text{ins}=m_\text{sph}/T$. Reduction of the large tunneling action
$S_\text{ins}$ by $\approx 10\%$ indicates an enhancement for the processes accociated with the energon, which is described by a large exponential factor 
\begin{equation}
\mathcal{K}\,=\,\exp(\,0.104 \,S_\text{ins})~.
\label{Ken}
\end{equation}
Take, for example,  the electro-weak theory. Then the instanton action is very large,  $S_\text{ins}=2\pi/\alpha$, $\alpha=g^2/4\pi\approx1/137$. This fact makes the instanton tunneling process strongly forbidden, though, as t'Hooft mentioned \cite{PhysRevLett.37.8}, it presents interest since it results in the barion non-conservation.
Equation (\ref{Ken}) shows that the enhancement to the tunneling process at finite temperatures due to the energon can be as large as $\mathcal{K}\approx 10^{39}$, which is a huge boost, though the remaining $90\%$ of the action still represents a formidable suppression. For QCD processes the enhancement provided by the energon is less dramatic, but the instanton is a more probable phenomenon there.
 
\section{Conclusion}
\label{Conclusion}

We discussed a family of configurations of the Yang-Mills  field, referring to it as the energon, which exhibits several interesting properties. It can be defined via the minimum of the classical action, and hence it is a solution of classical equations of motion, albeit this minimum and the corresponding equations are taken with restrictions, which state that the Euclidean energy and size of the energon are presumed fixed and play the role of parameters. Another way to define the energon is through a relation between  its electric and magnetic fields. These fields are collinear, and their ratio is a function of the Euclidean time, which is specified by the energy conservation law. The instanton and COS-sphaleron prove to be particular examples of the energon. They appear when the energon energy takes on specific values, i. e. zero and (minus) sphaleron mass for the instanton and sphaleron respectively. The energon possesses nonzero Chern-Simons number. Due to this reason in Minkowsky space it can be considered as a process, which provides transition through the effective barrier, which separates valleys with different topological numbers. 
We found that for temperatures in the vicinity of $\rho T\simeq 3/8$, the energon strongly, exponentially enhances the probability of this transition. 

Several problems arise for further consideration. The energon was derived here by a deformation of the single-instanton solution. It seems plausible that multi-instanton solutions can also be deformed in a similar way, leading to multi-energon configurations. If this anticipation proves true, the result may shed new light on the properties of multi-sphaleron configurations. An important issue is the interaction of the energon with other fields, fermions and scalars. It is known that the instanton interacts with  them very strongly, revealing nontrivial features related to the anomaly \cite{'tHooft:1976fv}. 
The sphaleron also shows very strong interaction with other fields \cite{Klinkhamer:1984di,Shuryak:2003xz}.
It is inevitable therefore that the energon should interact with other fields strongly.
This problem is directly related to possible practical applications, in which one has to specify whether the energon takes place in the Higgs phase of, for example, electroweak theory, or in the confining phase, as for the case of QCD. The interaction of the energon with external fields in these phases may lead to different results. A related issue presents connection between the energon and sphaleron solution of \cite{Klinkhamer:1984di}. This particular sphaleron incorporates self-consistently strong interaction of the gauge and Higgs fields. Therefore  in order to search for a possible connection between the energon and the sphaleron of \cite{Klinkhamer:1984di} one has to establish how the energon interacts with the scalar field.

\appendix
\section{Gauge transformations}
\label{gauge}
Take the matrix form for potentials and fields of the SU(2) gauge theory 
\begin{align}
&A_\mu~\,=~\frac{1}{2i}\,A_\mu^a\tau^a~,
\label{A2i}
\\
&F_{\mu\nu}\,=\,\frac{1}{2i}\,F_{\mu\nu}^a\tau^a~.
\label{F2i}
\end{align}
A gauge transformation induced by the matrix $U\,\in \,\text{SU(2)}$ in this notation reads
\begin{align}
&A_\mu\rightarrow A_\mu^\prime\,=\,U^{-1}A_\mu U+U^{-1}\partial_\mu U~,
\label{gt}
\\
&F_{\mu\nu}\rightarrow F_{\mu\nu}^\prime\,=\,U^{-1}F_{\mu\nu} U~.
\label{Fgt}
\end{align} 
Consider the potential $A_\mu=(\boldsymbol{A},A_4)$, which has the following specific form
\begin{align}
&\boldsymbol{A}\,=\,\frac{1}{2i}\,\Big[~\big( \boldsymbol{\tau}-\boldsymbol{n}\,(\boldsymbol{n}\cdot \boldsymbol{\tau})\big)\,
\frac{f \,\cos \alpha}{r} 
\nonumber
\\ 
&\quad\quad\quad\quad+\,\boldsymbol{n} \times \boldsymbol{\tau}\,\frac{1+f \,\sin \alpha}{r}\,+
g~ \boldsymbol{n} \,(\boldsymbol{n}\cdot \boldsymbol{\tau})
\,\Big]~,
\label{Am}
\\
&
A_4\,=\,\frac{1}{2i}\,h~(\boldsymbol{n}\cdot \boldsymbol{\tau})~,
\label{A4}
\end{align}
where $\alpha,f,g$ and $h$ are some functions of coordinates, and $\boldsymbol{n}=\boldsymbol{r}/r$. As Witten noticed \cite{Witten-PhysRevLett.38.121} a gauge transformation of this potential with the matrix
\begin{equation}
U\,=\,\exp\Big(\,-\frac{i\,\gamma}{2}~\,\boldsymbol{n}\cdot \boldsymbol{\tau}~\Big)~,
\label{Ug}
\end{equation}
where $\gamma$ is a function of $r$ and $\tau$, $\gamma=\gamma(r,\tau)$, results in the potential $A_\mu^\prime$ expressed via the same Eqs.(\ref{Am}),(\ref{A4}), in which parameters $\alpha,f,g,h$ are modified, the gauge transformation  $A_\mu\rightarrow A_\mu^\prime$ results in modification $\alpha,f,g,h \rightarrow \alpha^\prime,f^\prime,g^\prime,h^\prime$, where
\begin{align}
&\alpha^\prime\,=\,\alpha+\gamma~,
\label{alpha'}
\\
&f^\prime\,=\,f~,
\\
&g^\prime\,=\,g+\partial_r\gamma~,
\\
&h^\prime\,=\,h+\partial_\tau\gamma~.
\label{h'}
\end{align}
Write now the topological current (\ref{Kmu}) in the notation of (\ref{A2i}),(\ref{F2i})
\begin{equation}
K_\mu\,=\,-\frac{\epsilon_{\mu\nu\lambda\sigma} }{16\pi^2}~
\text{Tr}\Big( \,F_{\nu\lambda}A_\sigma-\frac{2}{3}\,A_\nu A_\lambda A_\sigma\,\Big)~.
\label{Kmu2i}
\end{equation}
Straightforward calculation shows, see e. g. discussion in \cite{Weinberg:1996kr}, that under a general gauge transformation (\ref{gt}),(\ref{Fgt}) the current exhibits variation $K_\mu\rightarrow K_\mu^\prime$, where
\begin{align}
&K_\mu^\prime\,=~K_\mu+J_\mu-\frac{\epsilon_{\mu\nu\lambda\sigma} }{8\pi^2}~
\partial_\nu\,\text{Tr}\big(\,\bar U\,A_\lambda\partial_\sigma U \,\big)~,
\label{KK'}
\\
&J_\mu~=~\frac{\epsilon_{\mu\nu\lambda\sigma} }{24\pi^2}~
\text{Tr}\,\big(\,\bar U\,\partial_\nu U\,\bar U\,\partial_\lambda U\,\bar U\,\partial_\sigma U \,\big)~.
\label{Jmu}
\end{align}
Here $\bar U\equiv U^{-1}$. 
The gauge dependence of the current shows that
the Chern-Simons number,  which is expressed via the fourth 
component (Euclidean notation) of the current density $n=\int K_4\,d^3r$, 
is influenced by the gauge transformation as well. For its variation, $n\rightarrow n^{\,\prime}$, one finds from (\ref{KK'}),(\ref{Jmu})  
\begin{equation}
n^{\prime}\,=\,n-\nu~.
\label{CS'}
\end{equation} 
Here $\nu$ is the Maurer Cartan form (which is applied here for the 3D-space and which equals the winding number, see \cite{Weinberg:1996kr})
\begin{equation}
\nu\,=\,\frac{\epsilon_{n l s} }{24\pi^2}~
\int\,\text{Tr}\,\big(\,\bar U\,\partial_n U\,\bar U\,\partial_l U\,\bar U\,\partial_s U \,\big)\,d^3r~.
\label{CM}
\end{equation}
It is taken into account in Eq.(\ref{CS'}) that the last, third term from Eq.(\ref{KK'}) 
is a derivative, and hence it does not contribute to $n^{\prime}$, the sign minus in Eq.(\ref{CS'}) complies with the Euclidean notation, in which $\epsilon_{1234}=1$.

Let us apply Eqs.(\ref{KK'}),(\ref{Jmu}), and ((\ref{CM}) for particular gauge transformations specified by Eq.(\ref{Ug}). In this case one finds
\begin{align}
\bar U \,\frac{\partial U}{\partial \mathbf{r} }\,=\,\frac{1}{2i}\,\Big[
&\big( \boldsymbol{\tau}-\boldsymbol{n}\,(\boldsymbol{n}\cdot \boldsymbol{\tau})\big)
\,\frac{\sin \gamma}{r}
\label{UdU}
\\
+ 
~&\boldsymbol{n} \times \boldsymbol{\tau}\,\frac{1-\cos \gamma}{r} +
\boldsymbol{n} \,(\boldsymbol{n}\cdot \boldsymbol{\tau})\,\frac{\partial \gamma}{\partial r}
\,\Big]~,
\nonumber
\end{align} 
derives then that
\begin{equation}
\epsilon_{n l s}
\text{Tr}\big(\bar U\,\partial_n U\,\bar U\,\partial_l U\,\bar U\,\partial_s U \big)
=-3\frac{1-\cos\gamma}{r^2}\,\frac{\partial \gamma}{\partial r}\,,
\label{nls}
\end{equation}
and finally obtains the following  representation for the Maurer Cartan form 
\begin{align}
\nu\,=\,-\frac{1}{24\pi^2}\,\int_0^\infty
\frac{3}{r^2}\,&(1-\cos\gamma)\,\frac{\partial \gamma}{\partial r} \,4\pi \,r^2\,dr
\nonumber
\\
=~-&\frac{1}{2\pi}\,\int_{\gamma_0} ^{\gamma_\infty}
(1-\cos\gamma)\,d\gamma~.
\label{nu}
\end{align}
Here $\gamma_{\,0}\equiv\gamma(r=0,\tau)$, and $\gamma_\infty\equiv\gamma(r=\infty,\tau)$.
It can be shown that for nonsingular gauge transformations $\gamma_{\,0}=p\pi$ and $\gamma_\infty=q\pi$, where $p,q$ are integers for which $p-q$ is even, so that $\nu=(q-p)/2$ is also an integer, as it should be in conventional cases, see \cite{Weinberg:1996kr}. 

There is though an important exception. In the gauge adopted in Eqs.(\ref{Aam}),(\ref{Aa4}) the potentials of the energon show singular behavior at spatial infinity, where they fall off too slowly $\propto 1/r$. To eliminate this shortcoming one has to fulfill the  gauge transformation, which inevitably absorbs the singular behavior at infinity as its own essential property. The latter manifests itself via the unusual boundary conditions $\gamma_{\,0}=0$, $\gamma_\infty=\pi$, compare Eq.(\ref{n}), in which $\pi$ appears with an odd coefficient 1. The result of this singularity manifests itself through  Eq.(\ref{nu}), which for this particular gauge transformation produces a semi-integer number
\begin{equation}
\nu\,=\,-\,1/2~.
\label{1/2}
\end{equation}  

\section*{Acknowledgment} 
I would like to thank Victor Flambaum and Edward Shuryak for discussions. A hospitality of staff and financial support of the Institute of Advanced Studies of the Massey University, where a part of this work was fulfilled, is acknowledged. The work is supported by the Austra1lian Research Council.

\end{document}